\documentclass[sigconf]{acmart}
\pdfoutput=1
\usepackage[utf8]{inputenc}

\usepackage{algorithmic}
\usepackage{cleveref}
\usepackage{comment}
\usepackage{csvsimple}
\usepackage{lipsum}
\usepackage{listings}
\usepackage{graphicx}
\usepackage{siunitx}
\usepackage{subfigure}
\usepackage{textcomp}
\usepackage{todonotes}
\usepackage{xcolor}
\usepackage{xspace}

\theoremstyle{definition}

\crefname{researchquestion}{RQ}{RQs}

\def\DataIndexedPapers{44581}
\def\DataIndexedPapersYearMin{1971}
\def\DataIndexedPapersYearMax{2020}
\def\DataIndexedPapersYearRange{1971--2020}
\def\DataIndexedPapersYearRangeLen{50}
\def\DataAllPapers{46585}

\def\DataAllPapersYearRange{1971--2020}
\def\DataMissingPapers{2004}
\def\DataMissingPapersPct{4.3}
\def\DataVenues{34}

\def\DataNgrams{577276382}

    \def\DataDoisSwengNotGi{14668}  \def\DataDoisSwengNotGiPct{36.2}      \def\DataDbSizeDefault{58\,GiB}  \def\DataDbSizeWIndexes{81\,GiB}

\def\DataDumpFilename{swepapers.pgsql.gz}

\def\NoDoi{\num{325}\xspace}  

\def\Notdownloaded{\num{1521}\xspace}

\def\NotParsedGrobid{483}  \def\ParsedGrobid{44581}  

 \copyrightyear{2022}
\acmYear{2022}
\setcopyright{acmlicensed}
\acmConference[MSR '22]{19th International Conference on Mining Software Repositories}{May 23--24, 2022}{Pittsburgh, PA, USA}
\acmBooktitle{19th International Conference on Mining Software Repositories (MSR '22), May 23--24, 2022, Pittsburgh, PA, USA}
\acmPrice{15.00}
\acmDOI{10.1145/3524842.3528494}
\acmISBN{978-1-4503-9303-4/22/05}

\title{The General Index of Software Engineering Papers}
\author{Zeinab Abou Khalil}
\email{zeinab.abou-khalil@inria.fr}
\orcid{0000-0003-0725-1024}
\affiliation{\institution{Inria}
  \city{Paris}
  \country{France}
}

\author{Stefano Zacchiroli}
\email{stefano.zacchiroli@telecom-paris.fr}
\orcid{0000-0002-4576-136X}
\affiliation{\institution{LTCI, Télécom Paris, Institut Polytechnique de Paris}
  \city{Paris}
  \country{France}
}

\begin{abstract}

  We introduce the \emph{General Index of Software Engineering Papers}, a
  dataset of fulltext-indexed papers from the most prominent scientific venues
  in the field of Software Engineering.  The dataset includes both complete
  bibliographic information and indexed n-grams (sequence of contiguous words
  after removal of stopwords and non-words, for a total of \num{\DataNgrams}
  unique n-grams in this release) with length 1 to 5 for
  \num{\DataIndexedPapers} papers retrieved from \DataVenues{} venues over the
  \DataAllPapersYearRange{} period.

  The dataset serves use cases in the field of meta-research, allowing to
  introspect the output of software engineering research even when access to
  papers or scholarly search engines is not possible (e.g., due to contractual
  reasons). The dataset also contributes to making such analyses reproducible
  and independently verifiable, as opposed to what happens when they are
  conducted using 3rd-party and non-open scholarly indexing services.

  The dataset is available as a portable Postgres database dump and released as
  open data.

\end{abstract}

\keywords{dataset, academic publishing, software engineering, ngrams,
  meta-analysis, natural language processing, fulltext index}

\begin{document}
\maketitle

\section{Introduction}
\label{sec:intro}

\emph{Meta-research}~\cite{ioannidis2010metaresearch,
  ioannidis2015metaresearch} is the use of the scientific method to study
\emph{science itself}. Meta-research is usually conducted by proxy, studying
the artifacts that science produces as byproducts like research papers,
datasets, reusable tools, etc. Meta-research studies are common in software
engineering too, primarily in the form of \emph{systematic literature reviews}
(for example~\cite{farias2016msrmapping, sun2016topicmodels}), which are the
recommended evidence-based tool of introspection in this
field~\cite{kitchenham2004ebse, kitchenham2009slrslr}, but also via analyses of
publication trends either in the field at large or in specific
venues~\cite{demeyer2013msrtrends, menzies2018swetrends,
  sahito2019swengpubtrends, gurcan2020distancelearningtrends}. While not yet
significantly practiced for this reason in software engineering, scientific
paper analyses are gaining traction in other fields, and most notably health
sciences, to uncover biases related to research
funding~\cite{alperin2019public}, gender diversity~\cite{andersen2020women},
grant evaluation~\cite{pina2021mariecurie}, publishing
policies~\cite{huang2020openaccess}, and more.

A significant burden for conducting meta-research on scientific papers is that
most scientific literature is not (yet) available as Open
Access~\cite{suber2003openaccess}. Most papers are accessible only via paywalls
that not all scholars have gratis access to, depending on the agreements that
their institutions have with scientific publishers; and that already excludes
all non-affiliated scholars who cannot afford to pay on their own for accessing
the papers they desire to study.

\emph{Reproducibility} is also a concern for meta-analyses. When conducted
using primarily 3rd-party and non-open scholarly indexing services (e.g.,
Google Scholar), both meta-analyses and simple scholarly searches are neither
reproducible by peers nor independently verifiable by other means. Search
results can change without notice, be biased by service providers, or outright
forbidden due to Terms-of-Service contractual provisions.

While these problem cannot be fixed without changing the balance of power in
scientific publishing, the availability of \emph{curated, open data paper
  indexes} can mitigate them. A recent initiative by Malamud, called the
\emph{General Index}~\cite{generalindex2021naturenews}, goes in this
direction. Malamud has indexed the full text of 107 million papers from major
journal across all scientific fields, extracting from them individual words and
n-grams (corresponding to phrases that are short enough for not being
considered copyrightable), and released the obtained index as open
data.\footnote{\url{https://archive.org/details/GeneralIndex}, accessed
  2021-12-15} The General Index allows scholars to search papers relevant for
their meta-analyses without having to rely on third parties like Google Scholar
or publisher websites. From there on, scholars who do have access to paywalls
can retrieve them without fees; others can already analyze paper metadata
(which are usually available as open data) or decide which paper to buy access
for depending on search results.
The coverage of software engineering in the General Index is limited though: we
have verified (see \Cref{sec:related} for details) that only major journals are
included, leaving out conference proceedings which are very important in the
field of software engineering.

\paragraph{Contributions.}

We introduce the \emph{general index of software engineering papers}, an
\textbf{open dataset of fulltext-indexed software engineering papers}, covering
\DataVenues{} major journals and conferences in the field (see
\Cref{tab:venues}), for a total of \num{\DataIndexedPapers} papers published
during the last \DataIndexedPapersYearRangeLen{} years
(\DataIndexedPapersYearRange). Papers metadata have been retrieved from
DBLP~\cite{ley2002dblp}, the reference bibliographic database for computer
science, and are integrated into the dataset. Each paper in the dataset has
been retrieved in full, converted to text, and text indexed by n-grams with
length 1 to 5 (allowing to search papers by individual words up to short
phrases), after removing English stop words. A total of \num{\DataNgrams}
unique n-grams have been indexed and can be looked up in the dataset.

\paragraph{Data availability.}

The dataset is \textbf{released as open data and available from Zenodo} at
\url{https://zenodo.org/record/5902231}, together with a full replication
package to recreate it from scratch. It is provided as a Postgres database dump
that expands to a \DataDbSizeWIndexes{} database (including indexes), which is
easily exploitable on commodity hardware.

 \section{Methodology and Reproducibility}
\label{sec:methodology}

\begin{figure}
  \includegraphics[width=\columnwidth]{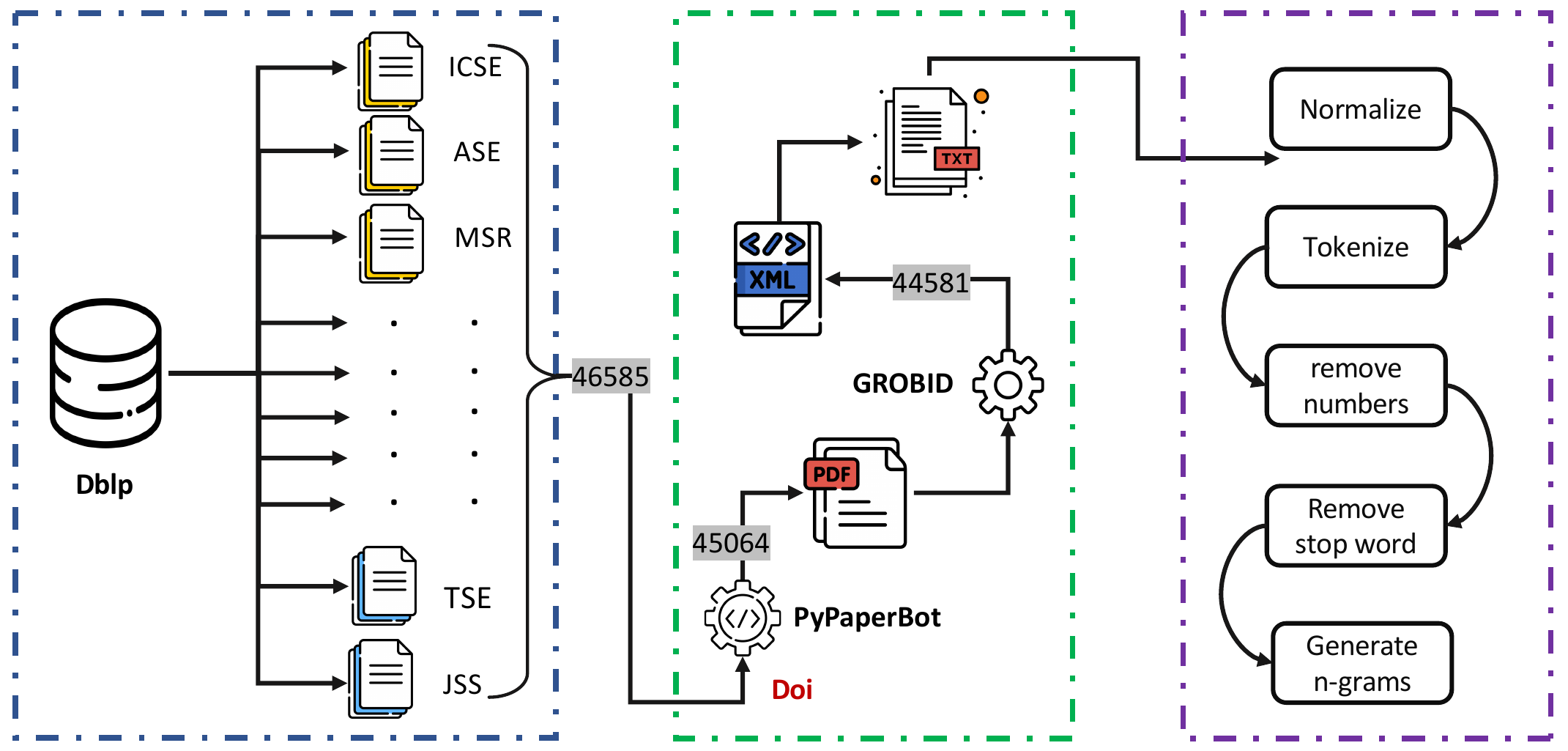}
  \caption{Dataset construction pipeline. }
  \label{fig:methodology}
\end{figure}

\Cref{fig:methodology} depicts the methodology used to assemble the dataset.

\paragraph{Data sources}

\begin{table}
  \caption{Breakdown of fulltext-indexed papers included in the dataset by
    venue, with coverage period for each.}
  \label{tab:venues}
  \footnotesize
\begin{tabular}{@{}l|p{4.4cm}|c|r@{}}
    \textbf{Acronym} & \multicolumn{1}{c|}{\textbf{Name}} & \textbf{Years} & \textbf{Papers} \\
    \hline
    \begin{filecontents*}{venues.csv}
acronym,name,years,papers
JSS,Elsevier - Journal of Systems and Software,1979--2020,4810
TSE,IEEE - Transactions on Software Engineering,1975--2020,3588
SPE,Software: Practice and Experience,1971--2020,3301
SW,IEEE Software,1984--2020,3260
ICSE,International Conference on Software Engineering,1988--2020,2947
IST,Information and Software Technology,1992--2020,2920
NOTES,ACM SIGSOFT Software Engineering Notes,1999--2020,2156
ASE,IEEE/ACM International Conference on Automated Software Engineering,1997--2020,2042
FSE,ACM SIGSOFT Symposium on the Foundations of Software Engineering,1993--2020,1795
ICSM,IEEE International Conference on Software Maintenance,1988--2013,1568
RE,IEEE International Requirements Engineering Conference,1993--2020,1321
IJSEKE,International Journal of Software Engineering and Knowledge Engineering,1999--2020,1151
ESE,Springer - Empirical Software Engineering,1996--2020,963
SOSYM,Software and System Modeling,2002--2020,890
MSR,Working Conference on Mining Software Repositories,2005--2020,808
ESEM,International Symposium on Empirical Software Engineering and Measurement,2007--2020,795
ICST,"IEEE International Conference on Software Testing, Verification and Validation",2008--2020,780
WCRE,Working Conference on Reverse Engineering,1993--2013,765
ISSTA,International Symposium on Software Testing and Analysis,1993--2020,746
SQJ,Software Quality Journal,1995--2020,721
MODELS,International Conference On Model Driven Engineering Languages And Systems,2005--2020,677
ICSME,International Conference on Software Maintenance and Evolution,2014--2020,629
ICPC,IEEE International Conference on Program Comprehension,2006--2020,626
FASE,Fundamental Approaches to Software Engineering,1998--2020,619
STVR,"Software Testing, Verification and Reliability",1991--2020,576
SMR,Journal of Software: Evolution and Process,2012--2020,576
SANER,"IEEE International Conference on Software Analysis, Evolution and Reengineering",2014--2020,544
TOSEM,ACM - Transactions on Software Engineering Methodology,1992--2020,513
ASEJ,Automated Software Engineering,1994--2020,507
REJ,Requirements Engineering Journal,1996--2020,504
SCAM,International Working Conference on Source Code Analysis \& Manipulation,2001--2020,467
GPCE,Generative Programming and Component Engineering,2002--2020,392
ISSE,Innovations in Systems and Software Engineering,2005--2020,381
SSBSE,International Symposium on Search Based Software Engineering,2011--2020,243
\end{filecontents*}
\csvreader[
      head to column names
    ]{venues.csv}{}{
      \acronym     & \name & \years & \num{\papers} \\
    }
    \\ \hline
    \textbf{Total} & &  \DataIndexedPapersYearMin--\DataIndexedPapersYearMax & \num{\DataIndexedPapers} \\
  \end{tabular}
\end{table}

For selecting the venues of \Cref{tab:venues} we relied on previous work,
adopting the list of Mathew et al.~\cite{menzies2018swetrends}.
We then obtained the list of all papers published in those venues using
DBLP~\cite{ley2002dblp}, the reference bibliographic database for computer
science publications, making its data available as XML
dumps.\footnote{\url{https://dblp.uni-trier.de/xml/}; specifically, we used the
  data dump \texttt{dblp-2021-04-01}.}

\paragraph{Data gathering}

For each selected paper, we retrieved complete bibliographic information from
the DBLP dataset, which includes DOIs (Digital Object Identifiers). 

Starting from DOIs, we then retrieved digital copies, in PDF format, of each
paper using the PyPaperBot paper retrieval
tool.\footnote{\url{https://github.com/ferru97/PyPaperBot}. We modified
  PyPaperBot to use DOIs as PDF filenames. The modified version is included in
  the replication package.} We excluded from download papers published in 2021,
as the year was not yet complete at the time of data gathering. Of the
remaining ones, \Notdownloaded papers could not be downloaded due to either
missing DOIs from DBLP (\NoDoi entries) or PyPaperBot failures to resolve DOIs
or retrieve the associated papers (1196 entries). This step took a few days for
PyPaperBot to download all papers.

\paragraph{Text indexing}

We converted PDF versions of the papers to plain text using
GROBID~\cite{GROBID} which, according to a recent
evaluation~\cite{tkaczyk2018machine}, is the best performing tools for
extracting content from scientific papers.
Out of the entire corpus, \num{\NotParsedGrobid} PDFs could not be parsed for
various reasons, such as different file format, corrupted PDFs, or PDFs
containing raster images instead of textual pages. In the end we obtained a
corpus of \num{\ParsedGrobid} text documents. The PDF conversion took
about a day to convert all the papers.

We cleaned up this textual corpus by applying standard NLP (Natural Language
Processing) preprocessing steps, as implemented by the Natural Language
Toolkit\footnote{\url{https://www.nltk.org/}, accessed 2022-01-07} (NLTK):
tokenization (splitting text into words, symbols, and punctuation), case
normalization to lower case, removal of non-alphabetic tokens, removal of
(English) stop words. Note that we applied neither stemming nor lemmatization,
as they can introduce clashes in search results for idiomatic expressions in
the domain, e.g., ``object orient\emph{ed}'', ``machine learn\emph{ing}'', etc.
If needed, stemming and lemmatization can be enforced at query time using
Postgres text search functions. 

Finally, we processed the (cleaned-up) fulltext of each paper to extract
$n$-grams with $n\in\{1,\ldots,5\}$, using NLTK. Extracted n-grams were
ingested into a Postgres database with the schema of \Cref{fig:schema},
together with the corresponding bibliographic information from the DBLP
dataset. This step took about 3 days on a standard issue work laptop.

 \section{Data Model}
\label{sec:datamodel}

\begin{figure}
  \includegraphics[width=\columnwidth]{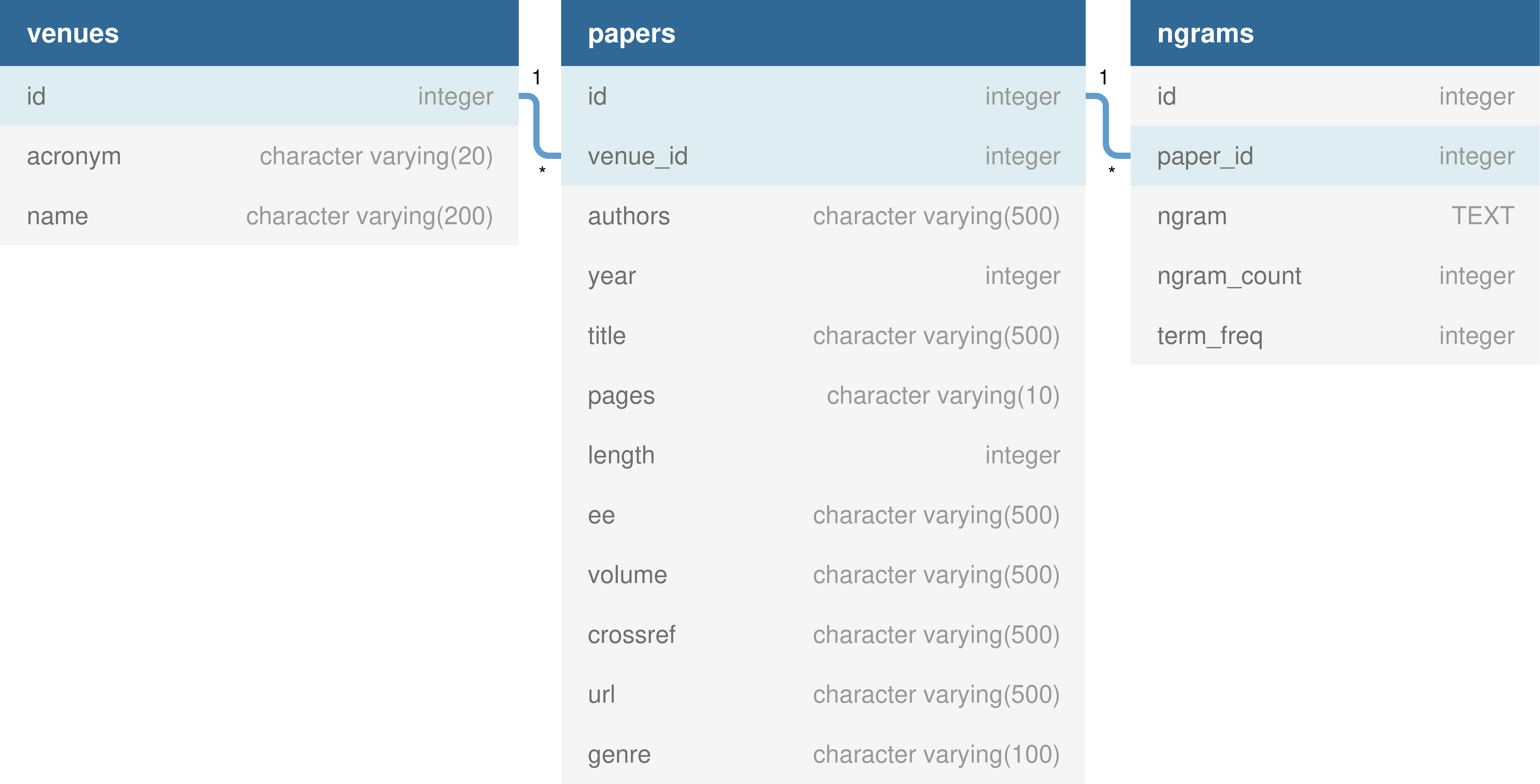}
  \caption{Relational database schema of the dataset.}
  \label{fig:schema}
\end{figure}

The dataset is available as a portable dump of a
Postgres~\cite{stonebraker1991postgres} database, which can be imported into
any instance of such a DBMS. The data model is straightforward for a textual
index and is shown in \Cref{fig:schema} as a relational database schema.

Querying will usually start from the \texttt{ngrams} table, which associates
n-grams (column \texttt{ngram}) to papers (\texttt{paper\_id}, referencing the
\texttt{papers} table). Each row also includes the length of the n-gram
(\texttt{ngram\_count}) and its frequency in the paper (\texttt{term\_freq}) as
a measure of relevance of the underlying concept in the study.

The \texttt{papers} table contains bibliographic metadata for papers in the
dataset. Most columns are self-descriptive, only a few caveats are worth noting:
\texttt{genre} denotes the paper venue using BiBTeX style names
(``inproceedings'' v. ``article'') inherited from DBLP; \texttt{ee} is a URL to
the electronic version of the paper; \texttt{length} is the paper length in
pages.
Most importantly, note that this
table contains information about \emph{all papers from selected venues}, even
those with no declared electronic version or with unparsable PDFs; to filter
the table on text-indexed papers join with the \texttt{ngrams} table (see
\Cref{sec:tutorial} for an example).

The \texttt{venue} table is linked from \texttt{papers} via the
\texttt{venue\_id} column and contains acronyms and full names of all venues in
the dataset.

 \section{Tutorial}
\label{sec:tutorial}

\subsection{Data import}

The dataset is available as a Postgres database dump which comes as a single
compressed file: \texttt{\DataDumpFilename}. Provided Postgres is installed and
the user has the needed permissions, the dataset can be imported as follows on
a UNIX shell:
\begin{lstlisting}[language=sh, escapechar=\%]
$ createdb swepapers
$ zcat \end{lstlisting}
The dataset can then be accessed using \texttt{psql swepapers}, or equivalent,
and perused via SQL queries.

By default, only indexes on primary key columns will be created. Most dataset
use cases will also need an index to efficiently lookup specific n-grams, which
can be created with the SQL command \lstinline[language=SQL]{create index on ngrams (ngram)}.
On disk the imported dataset occupies about \DataDbSizeDefault{} with default
indexes, \DataDbSizeWIndexes{} after adding the n-gram index.

\subsection{Sample queries}

The venue overview of \Cref{tab:venues} was obtained using query:
\begin{lstlisting}[language=SQL, basicstyle=\footnotesize\ttfamily]
select acronym, name, min(year), max(year),
  count(*) as papers
from venues
join papers on papers.venue_id = venues.id
where exists
  (select 1 from ngrams where paper_id = papers.id)
group by venues.id order by papers desc ;
\end{lstlisting}
note the filtering based on the \texttt{ngrams} table to skip non-text-indexed
papers; commenting it will instead give an overview of all papers whose
bibliographic information are available in the dataset.

\begin{table}
  \caption{Totals and ratios of software engineering papers mentioning
    ``machine learning'' (ML) over time (excerpt).}
  \label{tab:mlpapers}
\begin{tabular}{c|r|r|r}
    \textbf{Year} & \multicolumn{1}{c|}{\textbf{ML papers}} & \textbf{All papers} & \textbf{Ratio} \\
    \hline
    \begin{filecontents*}{ml_papers.csv}
year,sel,tot,pct
2009,93,1636,5
2010,87,1487,5
2011,121,1881,6
2012,157,1996,7
2013,223,2209,10
2014,203,2023,10
2015,246,1963,12
2016,309,2049,15
2017,350,2017,17
2018,459,2236,20
2019,547,2225,24
2020,705,2246,31
\end{filecontents*}
\csvreader[
      head to column names,
    ]{ml_papers.csv}{}{
      \year     & \sel & \tot & \pct\% \\
    }
  \end{tabular}
\end{table}

We can extract the total number and percentage of papers mentioning ``machine
learning'' (ML) over time as follows: 
\begin{lstlisting}[language=SQL,basicstyle=\footnotesize\ttfamily,morekeywords={with}]
with all_papers as (
  select year, count(*) as tot from papers group by year ),
sel_papers as (
  select year, count(*) as sel
  from papers join ngrams on ngrams.paper_id = papers.id
  where ngrams.ngram = 'machine learning'
  group by year )
select year, sel, tot, (sel * 100 / tot) as pct
from all_papers natural join sel_papers
order by year ;
\end{lstlisting} \Cref{tab:mlpapers} shows the query results for the past decade.

\begin{figure}
  \includegraphics[width=0.9\columnwidth]{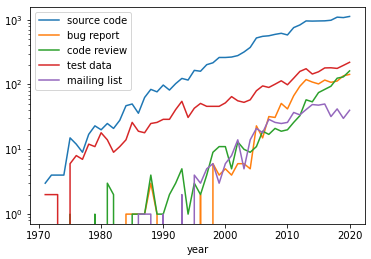}
  \caption{Number of software engineering papers mentioning selected types of
    software artifacts over time (log scale).
 }
  \label{fig:artifact-trends}
\end{figure}

Repeating similar queries with n-grams corresponding to various types of
software artifacts, one can study the evolution of the community interest in
them over time, as shown in \Cref{fig:artifact-trends} for selected artifacts.
A proper study of this topic will need to canvas relevant artifacts from the
dataset, determine associated n-grams, deal with their variants, weight term
frequency, etc.; this preliminary example is only meant to illustrate the
potential of the dataset.

\smallskip

\begin{figure}
  \includegraphics[width=0.9\columnwidth]{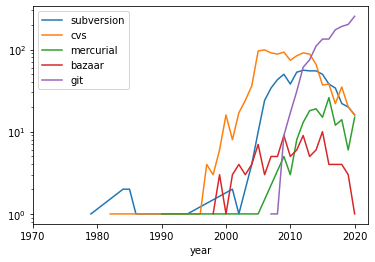}
  \caption{Number of software engineering papers mentioning selected version
    control systems (VCS) over time (log scale).}
  \label{fig:scm-trends}
\end{figure}

Software development is technology-driven and rapidly evolving; empirical
software engineering papers reflect that by looking into the usage of different
technologies over time. As an example, one can use the dataset to observe the
evolution of research interest in version control system (VCS) technology over
time as follows: 
\begin{lstlisting}[language=SQL,basicstyle=\footnotesize\ttfamily,morekeywords={with}]
with all_papers as (
  select year, ngrams.ngram 
  from papers join ngrams on ngrams.paper_id = papers.id
  where ngrams.ngram in 
  ('subversion', 'mercurial', 'cvs', 'git','bazaar'))
select year, ngram, count(ngram) as cnt
from all_papers 
group by year, ngram order by year ;
\end{lstlisting} \Cref{fig:scm-trends} shows the query results, highlighting Git as the top
mentioned VCS technology since the early 2010s, with every other VCS on the
decline.

\smallskip

To conclude, quiz time: what are the most common n-grams across software
engineering scientific literature? Queries like the following will satisfy your
curiosity (results excerpts are at the end of the paper, so that you can take a
guess!):
\begin{lstlisting}[language=SQL, basicstyle=\footnotesize\ttfamily]
select ngram, sum(term_freq) as freq
from ngrams join papers on ngrams.paper_id = papers.id
group by ngram order by freq desc ;
\end{lstlisting}

 \section{Limitations}
\label{sec:discussion}

\paragraph{Maintenance}

To remain valuable in the future, the dataset will need periodic releases and
curation. On the one hand, included venues will need to evolve as venues gain
prominence, split/merge, etc. Also, papers are published regularly and will
need to be indexed and added to the dataset. We plan to maintain the dataset
current, but we are also making it easier for \emph{others} to release new
versions of it in the future by (1) releasing the dataset creation and
maintenance tooling as part of the dataset, and (2) making the dataset update
incrementally: there is no need to re-index all previous papers to create a new
release, indexing new papers is enough and does not require significant
computing resources.

\paragraph{External validity}

The dataset is not meant to be representative of \emph{all} scientific venues
in software engineering, but only of the curated list of ``major'' ones
detailed in \Cref{tab:venues}. This curation criterion is admittedly arbitrary,
but is consistent with previous meta-analyses in the
field~\cite{vasilescu2013sweconf, menzies2018swetrends}. If other venues gain
prominence in the future, they can easily be added to future releases of the
dataset. Extending the dataset, both to include new venues and new articles, is trivial;
the process is detailed in the dataset documentation at
\url{https://zenodo.org/record/5902231}.

We rely on DBLP as ground truth for bibliographic information about papers
published in the selected venues, so we do not cover (nor could detect the lack
of) papers missing from DBLP. Given the preeminence of DBLP in computer science
and that of its dataset for meta-studies in the field, we consider this risk to
be very low.

\paragraph{Internal validity}

As detailed in \Cref{sec:methodology}, we could not obtain the fulltext of all
papers from selected venues, due to either lack of electronic edition/DOI
information in DBLP or unparsable PDFs. In the end, \DataMissingPapersPct\% of the papers in the
\DataAllPapersYearRange{} period (\DataMissingPapers{} out of
\num{\DataAllPapers}) could not be text indexed. As the amount is relatively
low and most impactful in the 1976--1989 period, we consider it to be an
acceptable limitation. Scholars of early software engineering history should,
however, consider manually retrieving and text indexing (e.g., using OCR
technology) the missing papers.

The choice of stopping at n-grams of length 5 is partly arbitrary and partly
dictated by copyright law. N-grams are an excerpt of works that are, for the most
part, copyrighted works owned by scientific publishers and released under
restrictive licensing terms. While there are no strict thresholds on the length
of sentences to be copyrightable, ``short'' excerpts are generally permitted
under fair use doctrine, whereas ``long'' excerpts will at some point violate
article licensing terms. Length 5 for n-grams is a very conservative choice,
shared by the General Index~\cite{generalindex2021naturenews}, which is a much
larger dataset in terms of indexed articles, currently considered acceptable in
terms of copyright law.

 \section{Related Work}
\label{sec:related}

The General Index (GI)~\cite{generalindex2021naturenews} is a fulltext-indexed
dataset of papers from scientific journals in all fields of science, and the
inspiration for this dataset. GI falls short of good coverage of software
engineering (SWE) though. Comparing the list of DOIs in the two datasets shows
that \num{\DataDoisSwengNotGi} from ours (\DataDoisSwengNotGiPct\%) are missing
from GI. Further inspection of the venues containing the word ``software'' in
GI shows that it does not index conference proceedings, explaining the gap.

Vasilescu et al.~\cite{vasilescu2013sweconf} published a historical dataset of
papers accepted at SWE conferences and associated program committee information
up to 2012. Kotti et al.~\cite{kotti2020msrdatapapers} investigated the use of
dataset papers published at MSR both using historical trends and interviewing
paper authors. Our dataset complements these works by adding the fulltext angle
of papers, and extending both the observation period and set of SWE venues.

We are not aware of related work other than the above providing relevant
datasets. On the other hand, a significant body of previous work has been
conducted using information analogous to those we provide with this dataset; in
particular: studies on trends in SWE in general~\cite{menzies2018swetrends,
  sahito2019swengpubtrends}, MSR~\cite{demeyer2013msrtrends,
  farias2016msrmapping}, topic modeling~\cite{sun2016topicmodels}, or distance
learning~\cite{gurcan2020distancelearningtrends}. In the field of SWE, this
kind of studies can be conducted or replicated in the future using this
dataset.

Several works~\cite{vine2006googlescholar, halevi2017googlescholar} in
bibliometry and medical sciences have pointed out the shortcomings and risks of
relying on Google Scholar as a bibliographic index and search engine. With this
dataset we are providing an independent and reproducible alternative for
performing fulltext searches across the SWE scientific literature.

 \section{Conclusion}
\label{sec:conclusion}

We introduced the General Index of Software Engineering Papers, a dataset of
\num{\DataIndexedPapers} papers from top venues in the field, fulltext-indexed
using n-grams. The dataset enables reproducible meta-analyses on software
engineering literature, without having to depend on third-party and non-open
scholarly indexing services.

As future work we plan to explore including in the dataset semantic-oriented
representations of the paper corpus such as latent semantic indexes and related
vectorial representations.

\smallskip \emph{Quiz answer:} the two most common n-grams in software
engineering literature are 1-grams ``software'' and ``data''; top 2-grams are
``test cases'', ``source code'', and ``software engineering''.


\end{document}